\documentclass{article}

\usepackage{graphicx}  
\usepackage{amsmath}   
\usepackage[compress]{cite}
\usepackage{amssymb}   
\usepackage{bm} 

\def\eq#1{{Eq.~(\ref{#1})}}

\def\frab#1#2{\left(\frac{#1}{#2}\right)}

\def\zpl{{zero-point-length}}

\def\ket#1{|#1\rangle}                    
\def\bra#1{\langle #1|}                   
\def\bk#1#2#3{{\langle #1|#2|#3\rangle}}  
\def\amp#1#2{\langle #1 | #2\rangle}      

\title{Probing the  Planck scale: The modification of the   time evolution operator  due to the quantum structure of spacetime}

\author{T. Padmanabhan\\
IUCAA, Post Bag 4, Ganeshkhind,
 Pune - 411 007, India.\\
email: paddy@iucaa.in}

\date{ }

\begin{document}

\maketitle

\begin{abstract}
The propagator which evolves  the wave-function  in NRQM, can be expressed as a matrix element of a time evolution operator: i.e $ G_{\rm NR}(x)= \langle{\bm{x}_2}|{U_{\rm NR}(t)}|{\bm{x}_1}\rangle$ in terms of the   orthonormal eigenkets $|{\bm{x}}\rangle$ of the position operator. In QFT, it is not possible to define a conceptually useful single-particle position operator or its eigenkets. It is also not possible to interpret the relativistic (Feynman) propagator $G_R(x)$ as evolving any kind of single-particle wave-functions. In spite of all these, it is indeed possible to express the  propagator of a free spinless particle, in QFT, as a matrix element $\langle{\bm{x}_2}|{U_{\rm R}(t)}|{\bm{x}_1}\rangle$ for a suitably defined time evolution operator and (non-orthonormal) kets $|{\bm{x}}\rangle$ labeled by spatial coordinates. At mesoscopic scales, which are close but not too close to Planck scale, one can incorporate quantum gravitational corrections to the propagator by introducing a zero-point-length. It turns out that even this QG corrected propagator can be expressed as  a matrix element $\langle{\bm{x}_2}|{U_{\rm QG}(t)}|{\bm{x}_1}\rangle$. I describe these results and explore several consequences. It turns out that the evolution operator $U_{\rm QG}(t)$ becomes non-unitary for sub-Planckian time intervals while remaining  unitary for  time interval is larger than Planck time. The results can be generalized to any ultrastatic curved spacetime.  
\end{abstract}

\section{Motivation}

\subsection{Propagators in NRQM and QFT}

    Consider a non-relativistic free particle with the Hamiltonian $H = \bm{p}^2/2m$. Its quantum dynamics can be completely characterized by the propagator\footnote{\textit{Notation:} I work in $1+3$ dimensions for definiteness, though the results can be trivially extended $1+d$ dimensions. Latin indices run over 0-3 while the Greek indices run over 1-3. I will use $x^i = (t, \bm{x})$ to denote the coordinates of an event even while discussing non-relativistic quantum mechanics (NRQM). The superscript $i$ etc. in $x_2^i, x_1^i$ will be often omitted and I will just write $x_2, x_1$ etc. for notational simplicity. The signature is mostly negative.}
    \begin{equation}
  G_{\rm NR}(x_2,x_1) = \theta(t)\frab{m}{2\pi it }^{n/2} \exp\frab{im|\bm{x}|^2}{2t};\qquad x\equiv x_2-x_1 
 \label{nrqmg}
    \end{equation} 
    The $\theta(t)$ in \eq{nrqmg} is somewhat conventional so that $G$ satisfies the equation $(i\partial_t - H) G_{\rm NR} = \delta_D(t)$ with a Dirac delta function on the right hand side. This factor is also  consistent with the feature that, when $G_{\rm NR}(x)$ is computed using a path integral, we only sum paths which go forward in time. But since non-relativistic Schrodinger equation is first order in the time derivative, one can use the same propagator --- without the $\theta(t)$ 
    factor --- to evolve the wave-function (either  forwards or) backwards in time; I will stick to the convention in \eq{one} to define $G_{\rm NR}$. (Nothing goes wrong in NRQM if the $\theta(t)$ is omitted.)
   
   For this propagator to consistently propagate the Schrodinger wave-functions, it must satisfy two crucial algebraic conditions: 
  \begin{equation}
  \lim_{t_b\to t_a} G_{\rm NR}(x_b, x_a)=\delta_D(\bm{x}_b - \bm{x}_a)
  \label{coin1}
 \end{equation} 
  \begin{equation}
  G_{\rm NR}(x_b, x_a) = \int d^n\bm{x}_1\ G_{\rm NR}(x_b, x_1)\,G_{\rm NR}(x_1, x_a)
  \label{complaw}
 \end{equation}
 One can directly verify from the explicit form of \eq{nrqmg} that these conditions do hold.
    The second condition \eq{complaw}, viz. the transitivity, is a strong constraint and is closely related to the fact that both wave-functions, and the propagator, satisfy a differential equation which is first order in time.  
    
   The NRQM propagator can be related to the Hamiltonian\footnote{This holds even for systems more general than free particle; but I will be only concerned with the free particle.} by expressing it as the matrix element of a time evolution operator in the form:
   \begin{equation}
     G_{\rm NR} (x) = \theta(t) \bk{\bm{x}_2}{U_{NR}(t)}{\bm{x}_1}\, ; \qquad  U_{NR}(t) = e^{-itH}
     \label{one}
    \end{equation} 
    where $x=x_2 - x_1$. Expressed in this form, the property in \eq{coin1} \textit{demands} the orthonormality of the kets: $\amp{\bm{x}}{\bm{y}} = \delta_D (\bm{x} - \bm{y})$ while the property in \eq{complaw} requires two conditions: (i) the completeness of the kets $\ket{\bm{x}}$ which allows the identity operator to be expressed as an integral over $d^3\bm{x} \, \ket{\bm{x}} \bra{\bm{x}}$ and (ii) the composition law for the evolution operator $U(t_1) U(t_2) = U( t_1 + t_2)$. 
    
    Let us move on from NRQM to the QFT of  a massive, free, spinless particle. In standard QFT, the (somewhat trivial) dynamics of the free field is entirely captured by the Feynman propagator $G_R (x) $ given by any one of these expressions:
    \begin{align}
     G_R(x) &= \frac{1}{i} \frac{1}{16\pi^2} \int_0^\infty\frac{ds}{s^2} \, \, \exp -i\left( \frac{x^2}{4 s} +  m^2 s\right)\label{grc}\\ 
    &=\int \frac{d^4p}{(2\pi)^4} \frac{ie^{-i px}}{p^2 - m^2+i\epsilon}\label{gra}\\ 
    & = \int \frac{d^3 \bm{p}}{(2\pi)^3 (2\omega_p)} \ e^{i\bm{p\cdot x} - i \omega_p |t|}\label{grb}
   \end{align} 
    Equation  (\ref{grc}) is the Schwinger's proper time representation of the propagator and is the most elegant way of describing it; this will be our work-horse in the later sections. Equation (\ref{gra}) is the more familiar expression for the Feynman propagator used in practical computations, which can be obtained by the 4-dimensional Fourier transform of \eq{grc} with respect to $x^i$. Similarly, \eq{grb} can be obtained  by a 3-dimensional Fourier transform of \eq{grc} or by a more familiar route of integrating over $p^0$ in \eq{gra} using standard contour integration techniques (See  section 1.4 of Ref. \cite{tpqft}). 
    
    Equation (\ref{grc}) and \eq{gra} are manifestly Lorentz invariant; one can  show \cite{tpqft} that \eq{grb} is also Lorentz invariant in spite of the occurrence of $|t|$.  
    To ensure convergence of the $s$ integral in \eq{grc}, we need to interpret $m^2$ as $m^2 - i \epsilon$ and $x^2$ as $x^2 - i \delta$. (Adding  a negative imaginary part to $m^2$ is a well known prescription. But note that, to ensure convergence near $s=0$, we  need to add a negative imaginary part to $x^2 $ as well. This is obvious when we consider the massless case and it ensures picking up the correct singular structure on the light cone.) I will not explicitly display $i \epsilon$ and 
    $i \delta$ except when it is relevant to the discussion.    Any of the integrals in \eq{grc}-\eq{grb} can be explicitly evaluated in terms of modified Bessel functions to give the result
    \begin{equation}
     G(x_2;x_1) = \frac{m}{4\pi^2 i \sqrt{x^2}} \, K_1(im\sqrt{x^2})
     \label{qft176}
    \end{equation} 
    We will not need this explicit form for most of our discussion.
    
    As an important  aside, let me stress that I have \textit{not} used the definition of propagator as the vacuum correlator of time ordered quantum fields. \textit{This is completely intentional.} In the later sections I will discuss the form of the propagator close to Planck scales. I want to work with a descriptor of the quantum dynamics  (of spinless particle of mass $m$) which is robust enough to survive (and be useful) close to Planck scales.  The propagator is a good choice for such a description because it is possible to define it  \textit{without using the notion of a local quantum field operator, commutation rules, vacuum state etc.}. In Appendix A, I mention three such definitions for the benefit of readers who tend to \textit{always} associate propagators with time-ordered correlators of quantum field.  None of the definitions in Appendix A use the formalism of a local field theory and its canonical quantisation, notions which may not survive close to Planck scales.
    
    In contrast to $G_{\rm NR}$, the relativistic propagator does \textit{not} satisfy the two conditions in \eq{coin1} and \eq{complaw}. 
    It satisfies a differential equation which is second order in time: $(\partial_t^2 - \nabla^2+m^2) G_R (x) = \delta_D (x)$.
    This is one of the key reasons why ideas like ``relativistic wave-functions'' involving single particle description run into serious conceptual difficulties. 
    
    \subsection{Mission Impossible?}\label{sec:mi}
    
    It will be interesting to ask: Can one find a representation for $G_R(x)$ which is similar in structure to that of $G_{\rm NR}$ in \eq{one}? That is, can we define some kets $\ket{\bm{x}}$, labeled by spatial coordinates and an operator $U_R(t)$, such that we can write 
    \begin{equation}
     G_R(x) = \bk{\bm{x}_2}{U_R(t)}{\bm{x}_1}
     \label{guess}
    \end{equation}
    At first sight, there are several obvious problems with a relation like \eq{guess}. 
    
    (i) The relativistic propagator $G_R$, unlike $G_{\rm NR}$, does not satisfy \eq{coin1} and \eq{complaw} and hence it is never going to propagate a Schrodinger-like wave-function. This, in turn, means that the kets $\ket{\bm{x}}$ cannot form an orthonormal set allowing a resolution of identity operator. 
    In fact, the 
    most crucial issue, in arriving at a relation of the form \eq{guess}, is  in the definition of the ket $\ket{\bm{x}}$ in quantum field theory. It is well known that defining a (particle) position operator and its eigenkets is conceptually dubious in  quantum field theory because particles cannot be localized.  That is, you cannot hope to define $\ket{\bm{x}}$ as an eigenket of a suitable position operator in QFT. They have to be defined by some indirect means and it is not clear whether such a definition will lead to a result like in \eq{guess}. 
    
    (ii) The non-relativistic propagator in \eq{one}, defined without $\theta(t)$ --- i.e., just as  the matrix element --- has the following property under time reversal: $G_{\rm NR} ( -t, \bm{x}) = G_{\rm NR} ^* (t,\bm{x})$;  time reversal leads to complex conjugation in NRQM. But the relativistic propagator in the left-hand-side  of \eq{guess} depends only on $t^2$ and hence is time-reversal invariant. This suggests that the evolution operator in \eq{guess} cannot have the standard form, viz., exponential of a Hermitian  operator which is linear in $t$. Therefore, we have no guarantee that the composition law $U(t_1) U(t_2) = U(t_1+t_2)$ will hold. 
    
    (iii) The left hand of \eq{guess} is Lorentz invariant. On the right hand side, space and time are clearly separated in the kets $\ket{\bm{x}_1}$, $\ket{\bm{x}_2}$ and in the operator $U(t)$. It is therefore not obvious how to find such a  structure which will be Lorentz invariant.

    The closest result to \eq{guess} one comes across in the literature is the following:
    The Schwinger representation for the propagator, in \eq{grc}, can also be expressed  as:
    \begin{equation}
     G(x) \propto \int_0^\infty ds\ \bk{x_2}{e^{-is\mathcal{H}}}{x_1} \, ; \qquad \mathcal{H}(p) \equiv -p^2 + m^2 - i\epsilon
     \label{grmatrix}
    \end{equation} 
    The \textit{integrand} looks similar to \eq{one} for $G_{\rm NR}$ but, of course, this is not in the form of \eq{guess} which I am seeking, because: 
    (a) The states $\ket{x_1}, \ket{x_2} $ are now labeled with the \textit{four} vectors $x^i$ rather than \textit{three} vectors $\bm{x}$ which I want in \eq{guess}. 
    (b) The (super) Hamiltonian $\mathcal{H} = -p^2 + m^2 - i\epsilon  = \Box + m^2 - i\epsilon$ is quite different from what we would expect for the relativistic particle $H(\bm{p}) = (\bm{p}^2 + m^2)^{1/2}$. 
    (c) Most crucially, we need to integrate over the Schwinger's proper time $s$ in \eq{grmatrix} in order to get the propagator; in \eq{guess} I want the propagator to be given directly as a matrix element.

    I will show, in the next section, that --- in spite of these issues --- one can indeed define the right hand side of \eq{guess} such that the equation holds!  Indirectly (but precisely) defined kets  $\ket{\bm{x}_1}, \ket{\bm{x}_2}$ along with an appropriate operator $U_R(t)$ is required for this job.  In fact, the result goes deeper.
    It has been suggested in several previous works \cite{zplimp, pid, zplextra, zplrecent, zpluse} that when the quantum gravitational corrections are taken into account, the propagator $G_R(x)$ gets modified with $x^2$ in \eq{grc} being replaced by $x^2 - L^2$ where $L^2 = \mathcal{O}(1)L_P^2=\mathcal{O}(1)(G\hbar/c^3)$ is the square of the  zero-point-length of the spacetime. 
    It turns out that one can modify the operator $U_R(t)$ such that an equation like \eq{guess} can actually lead to a propagator $U_{QG}(t)$ incorporating the zero-point-length. In fact, such a construction with quantum gravitational corrections actually explains some crucial features of the operator $U_R(t)$ which reproduces the standard propagator in QFT. In addition, $U_{QG}(t)$ gives us a glimpse of time evolution close to Planck scales.
    
  \section{Feynman propagator as a matrix element}
  
My aim is to \textit{define} the kets $\ket{\bm{x}}$ and the operator $U_R(t)$ such that \eq{guess} holds. I will first define the kets $\ket{\bm{x}}$ and then define the operator $U_R(t)$.  

Among the three issues (listed in the beginning of Sec. \ref{sec:mi} as (i),(ii) and (iii)) which one immediately notices  with \eq{guess}, the most important one is how to  define  $\ket{\bm{x}}$ without ever introducing a position operator for a particle. To do this, we will start with the eigenkets of the momentum operator and define $\ket{\bm{x}}$ using them. This can be done as follows.
 
A Hermitian  momentum operator  exists  in QFT as the generator of spatial translations in the one-particle sector of the standard Fock space. So, I will start by introducing a complete set of orthonormal momentum eigenkets, $\ket{\bm{p}}$ of this operator. We would then like $\amp{\bm{p}'}{\bm{p}}$ to be proportional to $\delta_D (\bm{p}- \bm{p}')$. This works  in NRQM but the integration over $d^3\bm{p}\delta_D (\bm{p}- \bm{p}')$ is not Lorentz invariant. The relativistically invariant measure for momentum integration is given by $d\Omega_{\bm{p}} \equiv d^3 \bm{p}/[(2\pi)^3 \Omega_{\bm{p}}]$ with $\Omega_{\bm{p}} = 2\omega_{\bm{p}}$. This requires us to  define the states $\ket{\bm{p}}$ with:
 \begin{equation}
 \amp{\bm{p}'}{\bm{p}} = (2\pi)^3 \Omega_{\bm{p}}\ \delta_D (\bm{p}- \bm{p}'); \qquad d\Omega_{\bm{p}} \equiv \frac{d^3 \bm{p}}{(2\pi)^3} \frac{1}{\Omega_{\bm{p}}}
 \label{four}
\end{equation} 
so that $\amp{\bm{p}'}{\bm{p}} d\Omega_{\bm{p}} = \delta_D (\bm{p}'- \bm{p}) d^3 \bm{p}$ and everything is Lorentz invariant.
 With this definition, the resolution of unity and the consistency condition on the momentum eigenkets, read as:
 \begin{equation}
1 \equiv \int d\Omega_{\bm{p}'} \ket{\bm{p}'}\bra{\bm{p}'};\qquad 
\ket{ \bm{p}} \equiv \int d\Omega_{\bm{p}'} \ket{\bm{p}'}\amp{\bm{p}'}{\bm{p}}
\label{fortynine}
\end{equation} 
 These relations can be taken care of by the choices in \eq{four}.
 In the integration measure as well as in the Dirac delta function, we have introduced a factor $\Omega_{\bm{p}}$ which, of course, cancels out in the right hand side of the second relation in \eq{fortynine}.

  I now  
  introduce the states $\ket{\bm{x}}$ labeled by the spatial coordinates. In NRQM they could be thought of as the eigenkets of the single-particle position operator $\hat{\bm{x}}(0)$. But, of course, in QFT, we do not have the natural notion of such a position operator; so I will not invoke such a conceptually dubious procedure. But there is a simple alternative: We can define $\ket{\bm{x}}$ by specifying its expansion  in terms of the basis vectors $\ket{\bm{p}}$. These expansion coefficients, in turn, can be chosen using the fact that 
  the momentum operator is the generator of spatial translations: So we will 
  \textit{define} $\ket{\bm{x}}$ by postulating the expansion coefficients for $\ket{\bm{x}}$ in the $\ket{\bm{p}}$ basis to be:
  \begin{equation}
\amp{\bm{p}}{\bm{x}} =  e^{-i\bm{x\cdot p}}                                           
\end{equation}
This is the same as the definition:
  \begin{equation}
 \ket{ \bm{x}} \equiv e^{-i\bm{x}\cdot \hat{\bm{p}}}|\bm{0}\rangle \equiv \int d\Omega_{ \bm{p}} e^{-i\bm{p\cdot x}}\ket{\bm{p}} ; \qquad  \amp{\bm{p}}{\bm{x}} =  e^{-i\bm{x\cdot p}}
 \label{fifty}
\end{equation}
We have set $\amp{\bm{p}}{\bm{0}}=1$ in the definition which, as it turns out, is the only consistent choice for Lorentz invariance. This defines $\ket{\bm{x}}$. 
  
Note that the kets $\ket{\bm{x}}$ etc. which we have defined, are \textit{not} orthogonal.  
  From the definition of $\ket{\bm{x}}$ in \eq{fifty}, it follows that:
 \begin{equation}
  \amp{\bm{y}}{\bm{x}} = \int d\Omega_{\bm{p}}  \, e^{-i\bm{p \cdot (x-y)}} \neq\delta_D(\bm{x-y})
  \label{fiftytwo1}
\end{equation} 
The evaluation of the integral leads to the standard result that $\amp{\bm{y}}{\bm{x}}$ decreases exponentially for separations larger than the Compton wavelength $\lambda_c\equiv (\hbar/mc)$. This is a direct consequence of the fact that particles cannot be sharply localized in QFT.

    Having defined the kets $\ket{\bm{x}}$ we now turn to the form of the operator $U_R(t)$ which will reproduce $G_R$ through \eq{guess}. The normal choice would have been $\exp[-it H(\bm{p})]$ with $H(\bm{p})\equiv (\bm{p}^2+m^2)^{1/2}$; this choice, however, will not lead to a $G_R$ through \eq{guess} because $G_R$ is an even function of $t$. To take care of it, I will define the operator $U_R(t)$ to be $\exp[-i |t| H(\bm{p})]$. (This form can also be `guessed' with a bit of reverse engineering from the structure of \eq{grb}.) 
    
    With these definitions of $\ket{\bm{x}}$ and $U_R(t)$, I claim that the relativistic  propagator is indeed given by the matrix element 
  \begin{equation}
   G_R(x) = \bk{\bm{x}_2}{U_R(t)}{\bm{x}_1}; \qquad       U_R(t)\equiv \exp[-i|t| H(\bm{p})]
   \label{claim}
  \end{equation} 
   The proof is straightforward. 
 Inserting a complete set of momentum eigenstates within the matrix element in \eq{claim}, and using the last relation in \eq{fifty}, we can evaluate the propagator explicitly to be:
 \begin{equation}
 G_R(x) = \bk{\bm{x}_b}{e^{-iH|t|}}{\bm{x}_a} = \int \frac{d^3 \bm{p}}{(2\pi)^3 (2\omega_p)} \ e^{i\bm{p\cdot x} - i \omega_p |t|}
 \label{fiftysix}
\end{equation} 
This gives the correct result for the propagator in the representation in \eq{grb}. The Lorentz invariance of \eq{claim} is assured because we know that the right-hand-side of \eq{fiftysix} is indeed Lorentz invariant, in spite of the appearance of $|t|$.

I will now provide an alternate derivation of the same result leading directly to the Schwinger's proper time representation in \eq{grc}. (This derivation has the advantage that it is easy to incorporate the zero-point-length, which I will do in the next section.)
 To do this, I  start with the easily proved (operator) identity:
\begin{equation}
 2H\int_0^\infty d\mu \, \exp\left( - i\mu^2 H^2 - \frac{i t^2}{4\mu^2}\right) =  \frab{\pi}{i}^{1/2}\,  e^{-i|t|H}
 \label{oneoneseven}
\end{equation} 
which allows us to  write, for $H^2=\bm{p}^2+m^2$, 
\begin{eqnarray}
 \bk{\bm{x}_b}{e^{-i|t|H}}{\bm{x}_a} &=&  \frab{i}{\pi}^{1/2}  \int_0^\infty d\mu \, e^{(-it^2/4\mu^2)}\ \bk{\bm{x}_b}{2H(\bm{p}) e^{-i\mu^2H^2(\bm{p})}}{\bm{x}_a}\nonumber\\
 &=&\frab{i}{\pi}^{1/2}  \int_0^\infty d\mu \, e^{(-it^2/4\mu^2)}\ e^{-i\mu^2m^2}\bk{\bm{x}_b}{2H(\bm{p}) e^{-i\mu^2\bm{p}^2}}{\bm{x}_a}
 \label{trick}
\end{eqnarray} 
The matrix element we need can now be evaluated by introducing a complete basis of momentum eigenkets $\ket{\bm{p}}$ with integration measure $d\Omega_p=d^3\bm{p}/[(2\pi)^32\omega_p]$ for the momentum integration. This gives,  with $\bm{x} \equiv \bm{x}_b - \bm{x}_a$ the result:
\begin{equation}
\bk{\bm{x}_b}{2H(\bm{p})e^{-i\mu^2\bm{p}^2}}{\bm{x}_a} =
  \int \frac{d^3\bm{p}}{(2\pi)^3}\frac{1}{2\omega_p} \, e^{i\bm{p}\cdot  \bm{x}} \, [2\omega_p e^{-i \mu^2 p^2}] = \frab{\pi}{i\mu^2}^{3/2} \frac{1}{8\pi^3}\exp \left(\frac{i\bm{x}^2}{4\mu^2}\right)
  \label{onetwozero}
\end{equation} 
The $2\omega_p$ arising from $2H$ in the left hand side of \eq{oneoneseven} cancels  the $(1/2\omega_p)$ in the measure of integration in the momentum space, giving a relatively simple result. Substituting \eq{onetwozero}
into \eq{trick} we leads to the final result, with $x^2 = x^ax_a=t^2  - \bm{x}^2$
\begin{eqnarray}
  \bk{\bm{x}_b}{e^{-i|t|H}}{\bm{x}_a} &=&  \frab{i}{\pi}^{1/2}  \frab{\pi}{i}^{3/2}\,\frac{1}{8\pi^3}\int_0^\infty \frac{ds}{2s^2} \, \exp\left( - \frac{ix^2}{4 s} - i m^2 s\right)\\
  &=& \frac{1}{i} \frac{1}{16\pi^2} \int_0^\infty\frac{ds}{s^2} \, \, \exp -i\left( \frac{x^2}{4 s} +  m^2 s\right)
  \label{oneoneight}
\end{eqnarray}
This is, of course, the Schwinger representation of the propagator in \eq{grc};  it is manifestly Lorentz invariant. 

The result in \eq{claim} is rather remarkable for several reasons. To begin with, the left hand side $G_R (x_2,x_1)$ is Lorentz invariant while in the right hand side, the matrix element, $\bk{\bm{x}_2}{U_{\rm R}(t)}{\bm{x}_1}$ separates space and time in a very concrete manner. Second, we do not have any simple physical \textit{interpretation} for the kets $\ket{\bm{x}}$ in QFT. Their \textit{definition}, through their expansion in the momentum basis, is rigorous and unambiguous  but it is not clear what they physically mean; this is again  because we do not have a notion of position operator. (In spite of several attempts in the literature, it has not been possible to define a conceptually sensible single particle position operator in QFT --- and there are excellent reasons for this failure; see e.g., \cite{qftnrqm}.) Third, 
the occurrence of $|t|$ in the evolution operator (and the propagator) is vital for the consistent interpretation of the theory with particles and antiparticles. (I will have more to say about this later on.) So the matrix element does not describe a single-particle propagation but actually encodes the sophisticated interplay of particle and antiparticle propagation in a rather succinct manner. Finally, I will show, --- in the next section --- that a similar  result holds even when we incorporate quantum gravitational corrections to the propagator through a \zpl\ in spacetime.

I will conclude this section by noting that there is a alternative integral representation of the evolution operator, using the function\footnote{This function was brought to my attention  by Karthik Rajeev, in the context of streamlining some discussion in \cite{krtp}.}
\begin{align}
	f[\nu,z]\equiv\int_{-\infty}^{\infty}\frac{ds}{(i\pi)}\left[\frac{s}{z^2-s^2-i\epsilon}\right]e^{-i\nu s}; \qquad (\nu>0)
	\label{deff}
\end{align}
defined in the entire complex plane with $z=x+iy$.
 This function is useful for defining the analytic continuation of $|t|$ when one proceeds from the Lorentzian to Euclidean sector with $t_E=it$. It is easy to verify that:
$f(\nu,z=x)=e^{-i\nu|x|}$ for  $\nu>0$ and $x$ along the real line. We also have
$f(\nu, z=iy)=e^{-\nu|y|}$ for $\nu>0$ and $y$  real which gives rigorous meaning to treating $e^{-\nu|t_E|}$ as the Euclidean extension of 
$e^{-i\nu|t|}$. (We will need this result later.)  This  leads to  an integral representation, for any positive definite Hamiltonian operator $H$:
\begin{align}
	U_R(t)=f[H,t]=e^{-iH|t|}=\int_{-\infty}^{\infty}\frac{ds}{(i\pi)}\left[\frac{s}{t^2-s^2-i\epsilon}\right]e^{-iH s}
	\label{defH}
\end{align}
which expresses the operator $e^{-iH|t|}$ in terms of the operator $e^{-iHs}$. This, in turn, provides a curious interpretation of the propagator.  Our result in \eq{defH} allows us to write the propagator as:
\begin{equation}
G_R(t,\bm{x}_2,\bm{x}_1)=  \bk{\bm{x}_2}{e^{-iH|t|}}{\bm{x}_1}=\int_{-\infty}^{\infty}d\tau\ A(t;\tau)\bk{\bm{x}_2}{e^{-iH\tau}}{\bm{x}_1}
\label{newint1}
\end{equation} 
with\footnote{This expression is superficially  similar to that in \eq{grmatrix} but, of course, is distinct from it. The kets in \eq{newint1} are labeled by spatial coordinates, $\bm{x}$, while the kets in \eq{grmatrix} are labeled by the spacetime coordinates $x^i$. The Hamiltonian in \eq{newint1} is just $H=(\bm{p}^2+m^2)^{1/2}$ while the (super) Hamiltonian in \eq{grmatrix} is $\mathcal{H}=-p^2+m^2$; and we do not have an amplitude like $A(t,\tau)$ appearing in \eq{grmatrix}.} 
\begin{equation}
 A(t;\tau)\equiv \frac{1}{(i\pi)}\left[\frac{\tau}{t^2-\tau^2-i\epsilon}\right]
\end{equation} 
In the integrand in the right hand side of \eq{newint1}, the factor $\bk{\bm{x}_2}{e^{-iH\tau}}{\bm{x}_1}$ gives the amplitude for propagation $\bm{x}_1 $to $\bm{x}_2$ in a (virtual) time interval of duration $\tau$; this is multiplied by the amplitude $A(t;\tau)$ for a virtual time interval $\tau$ to correspond to a physical time interval $t$. On integrating this expression over all values of virtual time interval $\tau$, we get the amplitude for propagation $\bm{x}_1$ to $\bm{x}_2$ in a physical time interval $t$. All the physics of particle-antiparticle propagation encoded in the $|t|$ factor of $\exp-iH|t|$ is eliminated by introducing a virtual  time interval and the amplitude $A(t;\tau)$. Instead of summing over virtual paths which go both forward and backward in time, we are summing over paths connecting the same $\bm{x}_1 $ and $\bm{x}_2$ but with different time intervals, ranging over the whole real line.\footnote{
I stress that we are doing standard QFT here. In fact, \eq{newint1} can be thought of as an integral convolution which converts the Wightman function 
$\bk{\bm{x}_2}{e^{-iH\tau}}{\bm{x}_1}=\bk{0}{\phi(\tau,\bm{x}_2)\phi(0,\bm{x}_1)}{0}$ 
to the Feynman propagator 
$\bk{\bm{x}_2}{e^{-iH|t|}}{\bm{x}_1}=\bk{0}{T[\phi(t,\bm{x}_2)\phi(0,\bm{x}_1)]}{0}$
}

\section{Propagator with quantum gravity  corrections}

 There exists a well-defined regime in which one can meaningfully talk about QG corrections to the standard QFT propagator. I will first describe this  context and then introduce the QG-corrected propagator. I will then show that the QG-corrected propagator can also be expressed as a matrix element, in the form of \eq{guess}, \textit{with the \textit{same} kets $\ket{\bm{x}}$} but with a modified evolution operator $U_{QG}(t)$. This, in turn, gives us some insight into time evolution close to Planck scales.

\subsection{Mesoscopic scales and the \zpl}

 I will consider a region of curved spacetime in which the curvature length\footnote{At any given event $\mathcal{P}$, the $L_{curv}$ could be defined in terms of typical curvature components; e.g., we can define $L_{curv}^{-2}=\sqrt{R^{abcd}R_{abcd}}$ evaluated at $\mathcal{P}$.} scale $L_{\rm curv}$ is much larger than Planck length: i.e., $L_{\rm curv} \gg L_P$. (If this condition is not satisfied we need the full machinery of QG which we do not have.) In that case, there exists a well-defined regime in which one can usefully introduce  QG corrections to the standard QFT propagator. 
 To do this,
  it is useful to introduce the notion of \textit{mesoscopic} regime, which interpolates between the \textit{macroscopic} regime (where one can use the standard formalism of QFT in CST) and the \textit{microscopic} regime, very close to and even smaller than the Planck scale (which requires a full quantum gravitational description). 
 This mesoscopic regime 
 is close, but not too close,  to the Planck scale so that we can still introduce  some kind of effective geometric description, while incorporating quantum gravitational effects to the leading order.  
 
 What happens to the QFT propagator at mesoscopic scales? The classical geometrical description will be modified close to Planck scales in a  manner which is at present unknown. However, we can  capture the most important effects of quantum gravity by introducing a \zpl\ to the spacetime \cite{pid, zplextra,zplrecent, zpluse}. This is based on the idea that the \textit{dominant} effect of quantum gravity at mesoscopic scales can be described by assuming that the path length $\sigma^2(x_2,x_1)$ in the \textit{Euclidean} sector\footnote{The \zpl\ is added to the spatial distance in the Lorentzian sector. With our signature, in flat spacetime, this involves the replacement of $x^2\equiv (t^2-\bm{x}^2)$ by $x^2-L^2=(t^2-\bm{x}^2-L^2)$ etc.}
has to be replaced by $\sigma^2(x_2,x_1) \to \sigma^2(x_2,x_1)+ L^2$ where  $L^2$ is of the order of Planck area $L_P^2 \equiv (G\hbar/c^3)$. 

It is possible to work out how this modification  translates to the form of the propagator. One can show that \cite{pid} the Euclidean propagator is now modified to: 
 \begin{equation}
  {G}_{QG}(x,y;m)=\int_0^\infty  ds\ e^{-m^2s-L^2/4s}K_{std}(s; x,y)
  \label{a15}
 \end{equation} 
 where $K_{\rm std}$ is the  zero-mass, Schwinger (heat) kernel given by $K_{\rm std} (x,y;s) \equiv \bk{x}{e^{s\Box_g}}{y}$. The $\Box_g$ is the Laplacian in the background space(time).
 Recall that  the leading order behaviour of the heat kernel is given by $K_{\rm std} \sim s^{-2}\exp[-\sigma^2(x,y)/4s]$ where $\sigma^2$ is the geodesic distance between the two events; therefore, the modification in \eq{a15} amounts to the replacement $\sigma^2 \to \sigma^2 +L^2$ to the leading order, which makes perfect sense.
 
Analytic continuation will give the propagator with \zpl\ in the Lorentzian sector. In the flat spacetime, we now get the propagator, incorporating  the \zpl\ of the spacetime to be:
 \begin{equation}
 G_{QG}(x)= \frac{1}{i} \frac{1}{16\pi^2} \int_0^\infty\frac{ds}{s^2} \, \, \exp -i\left( \frac{x^2-L^2}{4 s} +  m^2 s\right)
   \label{qgp1}
  \end{equation} 
which is manifestly Lorentz invariant. 
(To ensure convergence of the $s$ integral at the two limits, we must interpret $x^2$ as $x^2 - i\delta$ and $m^2$ as $m^2 - i\epsilon$. This, of course, was required even in the standard QFT propagator (with $L=0$) given by the Schwinger representation in \eq{grc}. No new regulator is needed due to the addition of \zpl.)

Before proceeding further,  let me stress some aspects of this approach --- which incorporates \zpl\ into the propagator --- for the sake of conceptual completeness. This will be helpful to readers (and referees!) who are not sufficiently familiar with previous work   on this approach.

\begin{itemize}
 
 \item Rigorously speaking, To study  the effects of quantum gravity at \textit{mesoscopic} scales we need  to start from a full theory of quantum gravity at \textit{microscopic} scales (for \textit{both} spacetime geometry \textit{and matter fields}) and work out a suitable coarse-grained approximation, involving an effective geometry, at mesoscopic scales. However, since we do not at present  possess the complete theory of QG or the description of matter fields at the microscopic scales, one needs to make some working hypothesis to proceed further. The idea of introducing the \zpl\ by the ansatz $\sigma^2 (x_2,x_1) \to \sigma^2(x_2,x_1) + L^2$ should be thought of as such a working \textit{hypothesis} which is postulated to make further progress. this  idea has been introduced and explored extensively in the past two decades or so in the  literature \cite{pid, zplextra, zplrecent, zpluse}. in this paper, I will be exploring some further consequences of this approach.

  \item By working directly with the propagator, we  bypass several nuances of standard QFT which may \textit{all} require some unknown form of revision at mesoscopic scales. In this approach we exploit the fact that both the dynamics \textit{and the symmetries} of a free quantum field, propagating in a curved geometry, is completely encoded in the Feynman propagator. So, if we have an ansatz to incorporate the  QG effects in the propagator, we obtain a \textit{direct} handle on both the dynamics and the symmetries of the theory at mesoscopic scales. This is an efficient procedure which encourages us to  work directly with the propagator containing QG corrections, without worrying about  the (unknown) modifications to the standard formalism of QFT at mesoscopic scales. 
  
  \item  One important consequence of working directly with the propagator is the following: The  diffeomorphism invariance in a curved geometry and --- as a special case -- Lorentz symmetry in flat spacetime is preserved in this approach. The prescription $\sigma^2 (x_2,x_1) \to \sigma^2(x_2,x_1) + L^2$ is generally covariant when $L$ is treated as a constant scalar number. In flat spacetime, this modification will replace $(x_2 - x_1)^2$ by $(x_2 - x_1)^2 + L^2$ which is clearly Lorentz invariant. (The mere introduction of a constant, scalar, length scale into the propagator will not violate Lorentz invariance, as should be obvious from the fact that the propagator for the massive scalar does depend on the length scale $m^{-1}$ and is still perfectly Lorentz invariant. The appearance of this length scale $L$ in the propagator has the same conceptual status as the appearance of, for example, the Compton wavelength $m^{-1}$ in the propagator.) It is not obvious how such symmetries can be preserved at the mesoscopic scales when we modify the formalism of QFT, with the usual canonical quantization, Fock basis etc. Using the propagator to encode the dynamics as well as the symmetries helps us to bypass such non-trivial issues; this is one reason why this formalism was powerful enough to do \textit{concrete} computations in a wide variety of contexts in the past literature. The results of such computations (see e.g., the extensive set of computations  ref. \cite{zpluse})  demonstrate the  general covariance and Lorentz invariance explicitly.\footnote{Some other prescriptions in the literature for introducing a `minimal length' do create problems for Lorentz invariance but \textit{our} prescription does not; it \textit{is} generally covariant.}
  
  \item   The propagator has a completely geometric interpretation in terms of a world line path integral --- see \eq{three} of Appendix A --- which does not use the formalism of fields, canonical quantisation etc. (This is outlined in Appendix A for the sake of those who \textit{always} think of a propagator as a two-point function of a field; you can define the propagator without introducing the notion of a field or its canonical quantisation.) Further, the action for a relativistic particle possesses a simple --- but not well-appreciated --- feature. The relativistic particle action is obviously expected to be a functional of the form $A[x^a(\tau); x_1,x_2]$; that is, it is a \textit{functional} of the world-line $x^a(\tau)$ and a \textit{function} of $x_2$ and $x_1$. But, it can be expressed  purely as a \textit{function} $A = A(\ell)$ of the length of the path $\ell[x^a(\tau); x_2,x_1]$, which carries the functional dependence on the world-line $x^a(\tau)$. This \textit{geometrical} structure of  action for the relativistic particle is a very special; in contrast, the standard action for the non-relativistic particle \textit{cannot} be expressed purely as a function of the length of the path. It is this feature which allows us to translate the modification of path lengths in spacetime (by the addition of the \zpl) to the modification of the relativistic action and thus of the propagator. This, in turn, allows us to preserve all the relevant symmetries of the theory and directly compute the corrections to the propagator at mesoscopic scale.

 \end{itemize}

After this aside, let me now return to the main topic. Once the Schwinger representation of the propagator is known, we can immediately write down the expression corresponding to \eq{grb}. One can, of course,  obtain it by Fourier transforming \eq{qgp1} with respect to the spatial coordinates $\bm{x}$. More simply, one can reason out as follows:
The equivalence of \eq{grb} with \eq{grc}  holds for any real parameter $t$. Therefore, replacing $|t|$ by $(t^2 - L^2)^{1/2}$ in \eq{grb} is equivalent to replacing $x^2 \equiv t^2-\bm{x}^2$ by $x^2 - L^2$ in \eq{grc}. But this is precisely the introduction of \zpl\ which converts $G_R(x)$ to the quantum corrected propagator $G_{QG}(x)$. Therefore, we  get the result:
\begin{equation}
G_{QG}(x)= \int \frac{d^3 \bm{p}}{(2\pi)^3 (2\omega_p)} \ e^{i\bm{p\cdot x} - i \omega_p \sqrt{t^2-L^2}}
\label{grb1}
\end{equation} 
which just involves replacing $|t|$ by $(t^2-L^2)^{1/2}$ in \eq{grb}. This expression is rather remarkable and we will exploit it in the next section.\footnote{The mesoscopic scale description is valid only when $t^2\gtrsim L^2$; in this range the phase remains real in \eq{grb1}. We will say more about this feature later on.}
These expressions \eq{qgp1}, \eq{grb1} describe the  QG corrections to the propagator  at mesoscopic scales.

To avoid possible confusion, let me mention the following (algebraic) fact: We all know that the measure $d^3 \bm{p}/ (2\omega_p)$ is Lorentz invariant; so is the standard combination $(\omega_p t-\bm{p\cdot x})$. It way appear, at first sight, rather surprising that the expression in the right hand side of \eq{grb1} is also Lorentz invariant --- which follows from the fact that the left hand side, which depends only on $x^2$ is Lorentz invariant --- in spite of $t$ being replaced by $(t^2-L^2)^{1/2}$. To understand this result, consider an arbitrary  scalar function $F$ of the Lorentz invariant variable $p^2-m^2$, say: $F(p^2-m^2)=F(\nu^2-\omega_{\bm{p}}^2)$ and its four-dimensional Fourier transform, written as:
\begin{equation}
I(x^2)\equiv \int d^4p F(p^2)e^{-ipx}=\int d^3 \bm{p}\ e^{i\bm{p\cdot x}}\int_{-\infty}^\infty d\nu\ F(\nu^2-\omega_{\bm{p}}^2)e^{-i\nu t}
\end{equation} 
The $\nu$ integration will lead to a function, say, $Q(t^2,\omega_{\bm{p}}^2)$ which can always be written as $Q=R(t^2,\omega_{\bm{p}}^2)/(2\omega_{\bm{p}})$ so that:
\begin{equation}
 I(x^2)=\int \frac{d^3 \bm{p}}{ (2\omega_{\bm{p}})}R(t^2,\omega_{\bm{p}}^2)e^{i\bm{p\cdot x}}
\label{ix2}
\end{equation} 
with
\begin{equation}
\frac{R(t^2,\omega_{\bm{p}}^2)}{(2\omega_p)}\equiv\int_{-\infty}^\infty d\nu\ F(\nu^2-\omega_{\bm{p}}^2)e^{-i\nu t};\qquad
F(\nu^2-\omega_{\bm{p}}^2)=\int_{-\infty}^\infty dt\ \frac{R(t^2,\omega_{\bm{p}}^2)}{(2\omega_p)}e^{i\nu t}
\end{equation} 
Clearly, the expression in the right hand side of \eq{ix2} is Lorentz invariant in spite of appearance. The expression in \eq{grb1} has exactly this form with $R=(2\pi)^{-3}e^{- i \omega_p \sqrt{t^2-L^2}}$. Its Fourier transform, $F(\nu^2-\omega_{\bm{p}}^2)$ can be expressed in terms of Bessel functions and its explicit form is given in the Appendix. A large class of integrals of the form $I(x^2)$ in \eq{ix2} can be Lorentz invariant without being \textit{manifestly} Lorentz invariant.

\subsection{Propagator with \zpl\ as a matrix element}

 The quantum corrected propagator, obtained by introducing a \zpl, is a rather strange beast. While it can be obtained from a path integral (see \cite{pid}  and Appendix A) and can be used to compute explicitly the QG corrections to several QFT/QED phenomena (see e.g, \cite{zpluse}), it \textit{cannot} be expressed as a time ordered correlator of a local quantum field. In the context of the current work, the  question arises as to whether  this propagator can also be expressed as a matrix element of some time evolution operator $U_{\rm QG}(t)$. If we could do that, it will throw some light into the concept of time evolution at mesoscopic scales close to Planck length. 
 
 I will now show that not only this can be done but also both the derivation and the result are extremely simple. I will show that all we need to do is to replace the time evolution operator $U_R=\exp(-iH|t_2-t_1|)$ by 
 \begin{equation}
  U_{\rm QG}(t) = \exp(-iH\sqrt{t^2-L^2}); \qquad t= t_2-t_1
 \end{equation} 
to get the correct result. I will first derive the result and discuss the implications afterwards.

 A  simple way to arrive at the correct answer is as follows: In \eq{fiftysix} the real variable $|t|$ goes for a ride on both sides of the equation. So if you replace $|t|$ by any other real variable, the equation will continue to hold. I will replace $|t|$ by $(t^2 - L^2)^{1/2}$ with the understanding that the positive square root is taken. This will lead to the result 
\begin{equation}
  \bk{\bm{x}_b}{e^{-iH\sqrt{t^2 - L^2}}}{\bm{x}_a} = \int \frac{d^3\ \bm{p}}{(2\pi)^3 (2\omega_p)} \ e^{i\bm{p\cdot x} - i \omega_p \sqrt{t^2 - L^2}}
 \label{grbqg1}
\end{equation} 
But the right hand side of \eq{grbqg1} is precisely the right hand side of \eq{grb1}. Therefore we immediately get the result:
  \begin{equation}
 G_{QG}(x) =\bk{\bm{x}_b}{e^{-iH\sqrt{t^2-L^2}}}{\bm{x}_a} 
 \label{final1}
\end{equation} 

  One can also obtain the same result from modifying the derivation leading to \eq{oneoneight}. 
In the operator identity in \eq{oneoneseven} the parameter $t$ goes for a ride on both sides; that is, the identity will hold with $t$ replaced by any other real quantity. I will replace $t^2$ in the left hand side by $(t^2-L^2)$ thereby getting the result:
\begin{equation}
 2H\int_0^\infty d\mu \, \exp\left( - i\mu^2 H^2 - \frac{i (t^2-L^2)}{4\mu^2}\right) =  \frab{\pi}{i}^{1/2}\,  e^{-iH\sqrt{t^2-L^2}}
 \label{new117}
\end{equation} 
That is all we need; it is obvious that the entire derivation proceeds exactly as before and leads to --- in place  of \eq{oneoneight} --- the modified result: 
\begin{eqnarray}
  \bk{\bm{x}_b}{e^{-iH\sqrt{t^2-L^2}}}{\bm{x}_a} 
  = \frac{1}{i} \frac{1}{16\pi^2} \int_0^\infty\frac{ds}{s^2} \, \, \exp -i\left( \frac{x^2-L^2}{4 s} +  m^2 s\right)
  \label{oneoneight1}
\end{eqnarray}
The right hand side, of course, is the QG corrected propagator so that we can now write:
\begin{equation}
 G_{QG}=\bk{\bm{x}_b}{e^{-iH\sqrt{t^2-L^2}}}{\bm{x}_a} 
\end{equation}
Since we expect the mesocopic scale description to be valid only for $t=t_2-t_1 >L$, the phase is real and the evolution operator is unitary for Hermitian $H$. I will now make a brief digression to show how these results can be generalized to a wider class of spacetimes and then discuss several implications of these results in Sec. \ref{sec:conclusion}.

\section{Aside: Generalization to ultrastatic spacetime}\label{sec:aside}

  The results in the previous two sections ---  related to the representation of $G_R$ and $G_{\rm QG}$ as matrix elements of the evolution operators --- remain valid in a wider class of curved spacetime (sometimes called ultrastatic) with the line element: 
  \begin{equation}
  ds^2 = dt^2 + h_{\alpha\beta}(\bm{x})\ dx^\alpha\, dx^\beta
  \end{equation} 
  (Note that, with our signature convention, $h_{\alpha\beta}$ will be a negative definite metric.) 
  The static nature of the spacetime ensures that both $G_{\rm R}(t,\bm{x}_2, \bm{x}_1)$  and $G_{\rm QG}(t,\bm{x}_2, \bm{x}_1)$ depends on time only through the difference $t\equiv (t_2-t_1)$. 
  I will first show that, in such a curved background, $G_{\rm QG}$ is obtained by replacing $t$ by $\sqrt{t^2 - L^2}$ in $G_R$. 
  I will obtain the expression for $G_{\rm QG}$ directly which will reveal this structure.  
  
  We start with the prescription for the propagator incorporating the \zpl\ in an arbitrary curved spacetime:
   \begin{equation}
  G_{\rm QG}(x_2,x_1) = \int_0^\infty ds\ e^{-im^2 s+(iL^2/4s)}\ \bk{x_2}{e^{-is\Box}}{x_1}
   \label{a2}
  \end{equation} 
  where the four-dimensional Laplacian $\Box$ separates into 
   \begin{equation}
 \Box = \frac{1}{\sqrt{-g}} \partial_a \left( \sqrt{-g}\, g^{ab}\partial_b\right) = \frac{\partial^2}{\partial t^2} + \frac{1}{\sqrt{h}}\partial_\alpha \left( \sqrt{h}\, h^{\alpha\beta} \partial_\beta\right) \equiv \frac{\partial^2}{\partial t^2} + \nabla_h^2
  \end{equation} 
  This guarantees that we can also separate the kets $\ket{x}$ into the direct product $\ket{t}\ket{\bm{x}}$ such that 
   \begin{equation}
  e^{-is\Box} \ket{x_1} = e^{-is\partial_t^2} \ket{t_1} \, e^{-is\nabla_h^2}\ket{\bm{x}_1}
  \end{equation} 
  We now introduce the eigenstates $\ket{\omega}$ of the one-dimensional operator $\partial_t^2$ and expand the kets $\ket{t_1}$ etc. in the form 
   \begin{equation}
 \ket{t_1} = \int_{-\infty}^\infty  \frac{d\omega}{2\pi} \, e^{i\omega t_1}\ket{\omega}\, ; \qquad \amp{\omega}{t} = e^{i\omega t}\, ; \qquad \amp{t}{\omega} = e^{-i\omega t}
  \end{equation}
and evaluate the time dependence of the matrix element as:
  \begin{equation} 
 \bk{x_2}{e^{-is\Box}}{x_1} = \int_{-\infty}^\infty \frac{d\omega}{2\pi}  e^{is\omega^2} e^{-i\omega t}  \bk{\bm{x}_2}{e^{-is\nabla_h^2}}{\bm{x}_1}
 =\left(\frac{i}{4\pi s}\right)^{1/2} e^{-(it^2/4s)} \bk{\bm{x}_2}{e^{-is\nabla_h^2}}{\bm{x}_1}
  \end{equation}
  Substituting this into \eq{a2}, we immediately find that $G_{\rm QG}$ depends on $t$ through the combination $(t^2 - L^2)$. Since $L=0$  reduces $G_{\rm QG}$ to $G_R$, we get the result we are seeking, viz.,
  \begin{equation}
 G_{\rm QG} (t, \bm{x}_2, \bm{x}_1) = G_{\rm R} \left(\sqrt{t^2 - L^2}, \bm{x}_2, \bm{x}_1\right) 
   \label{a8}
  \end{equation}
  This is, of course, completely analogous to what we found earlier in the special case of the flat spacetime. 
  
  I will next define a suitable set of kets $\ket{\bm{x}}$, labeled by the spatial coordinates, and prove that $G_R$ itself can be expressed as the matrix element 
  \begin{equation}
  G_{\rm R}(t, \bm{x}_2, \bm{x}_1) =  \bk{\bm{x}_2} {e^{-i|t|H}}{\bm{x}_1}
   \label{a9}
  \end{equation}
  where $H^2 = \bm{p}^2 + m^2$ with $\bm{p}^2$ evaluated using (negative of) the spatial metric $-h^{\alpha\beta}$. We can again  introduce the kets $\ket{\bm{x}}$  exactly as before, by using generalized mode functions in place of $e^{i\bm{p\cdot x}}$ which we used earlier. Let the eigenkets of the operator $H^2$ be $\ket{\omega,\mu}$ with 
  \begin{equation}
 H^2\ket{\omega,\mu} = \omega^2 \ket{\omega,\mu} 
  \end{equation}
 where $\mu$ collectively denotes all other parameters of the eigenket. (For example, in flat spacetime, we earlier labeled the eigenkets of $H$ by the three components of the momentum $\ket{\bm{p}}$ with $\omega^2=m^2+\bm{p}^2$. Instead, we could have traded off $p_x$ for $\omega$ and labeled the eigenkets by $\ket{\omega, p_y,p_z}$ so that $\mu=(p_y,p_z)$.)
 Further,  we can construct the propagator --- as a solution to $(\Box + m^2) G_R = \delta_D$ --- in terms of a complete set of orthonormal mode functions $F(x)$ which satisfy the homogeneous equation $(\Box + m^2)F =0$. In the ultrastatic spacetime, we can choose the mode functions to be $F=f_{\omega\mu}(\bm{x}) e^{\pm i\omega t}$, separating out the time dependence. We will choose $f_{\omega\mu}$ to be real, which can always be done,  for convenience. The relativistic propagator which satisfies the equation $(\Box + m^2) G_R = \delta_D$ can now be constructed in terms of the mode functions as: 
  \begin{equation}
  G_R(x) = \sum_{\omega,\mu} e^{-i\omega|t|} f_{\omega \mu} (\bm{x}_2) f_{\omega \mu} (\bm{x}_1) 
   \label{a10}
  \end{equation}
  We will now \textit{define} the kets $\ket{\bm{x}}$ by the expansion 
  \begin{equation}
  \ket{\bm{x}} =  \sum_{\omega,\mu} f_{\omega \mu} (\bm{x})\ket{\omega,\mu} \, ; \qquad \amp{\omega,\mu}{\bm{x}} =  f_{\omega \mu} (\bm{x})\, ; \qquad  
  \amp{\bm{x}}{\omega,\mu} =  f_{\omega \mu} (\bm{x})
  \end{equation}
  It follows that 
  \begin{equation}
 \bk{\bm{x}_2} {e^{-i|t|H}}{\bm{x}_1} =  \sum_{\omega,\mu} f_{\omega \mu} (\bm{x}_2)\, f_{\omega \mu} (\bm{x}_1)\, e^{-i\omega |t|}
  \end{equation}
  Comparing with \eq{a10}, we find that the right hand side is just $G_R$.  This immediately leads to the result quoted in \eq{a9}. 
  Combined with \eq{a8}, we find that the propagator incorporating the \zpl\ can again be expressed in the form 
  \begin{equation}
   G_{\rm QG} (t, \bm{x}_2, \bm{x}_1) = \bk{\bm{x}_2} {e^{-iH\sqrt{t^2 - L^2}}}{\bm{x}_1}
  \end{equation}
  in all ultrastatic spacetime. The results in the previous sections can be thought of as special cases when the spatial metric represents flat spacetime.

\section{Discussion}\label{sec:conclusion}

\subsection{Some consequences of the result}

The mesoscopic scale is defined to be close to  but somewhat larger than the Planck scale. This necessarily implies that the idea of a quantum corrected propagator is conceptually meaningful only if $t^2 > L_P^2 $ (and $|\bm{x}|^2 > L_P^2$). So, strictly speaking, our considerations in the last section is valid only when $(t^2-L_P^2) >0$. In that case, the phase of the modified evolution operator, $\exp(-iH \sqrt{t^2 - L_P^2})$ remains real and meaningful. It implies that we can talk about a unitary time evolution only when 
$t_2-t_1 >L_P$, which makes physical sense. The description in terms of a smooth geometry and a QG corrected propagator is conceptually dubious when the time interval $t_2 - t_1$ is sub-Planckian.

There are some interesting aspects of this (modified) time  evolution operator which is worth mentioning. We saw earlier that, in the standard QFT, the evolution operator is given by 
\begin{equation}
 e^{-i H|t|}=\theta(t)e^{-iHt}+\theta(-t)e^{iHt}; \qquad t=t_2-t_1
 \label{gtheta}
\end{equation} 
This shows that positive frequency modes are propagated forward in time while negative frequency modes are propagated backwards in time. This is also closely related to the notion of antiparticles and the propagator being a time-ordered product. All these becomes apparent (see e.g., \cite{tpqft}) when we look at a \textit{complex} scalar field for which the antiparticle is distinct from the particle. If we write the complex scalar field as the sum $\phi(x)\equiv A(x)+B^\dagger(x)$ with
\begin{equation}
A(x) \equiv \int d\, \Omega_\mathbf{p} A_\mathbf{p}e^{-ipx};\quad
B(x) \equiv \int d\, \Omega_\mathbf{p} B_\mathbf{p}e^{-ipx}
\end{equation}
where $A_\mathbf{p}$ and $B_\mathbf{p}$ are the standard annihilation operators,
then the propagator is given by:
\begin{equation}
 \bk{\bm{x}_2}{e^{-iH|t|}}{\bm{x}_1} = \theta(t)\bk{0} {A(x_2) A^\dagger(x_1)}{0} + \theta(-t) \bk{0} {B(x_1) B^\dagger(x_2)}{0}
\label{qft1134}
\end{equation} 
which clearly shows that the $|t|$ is vital to ensure proper propagation of particles and antiparticles. 

The following (algebraic) fact is equally important. The solutions of the Klein-Gordan equation will involve mode functions with time evolution $f\sim e^{\pm i\omega_{\bm{p}}t}$ without any $|t|$. The bilinear forms of mode functions used in constructing the propagator (which only  depends on $(t_2-t_1)$) can only involve the products like  $f(t_1)f^*(t_2)$ etc. will go as  $e^{\pm i\omega_{\bm{p}}(t_2-t_1)}$, again without $|t|$. To get the $|t|$ in the evolution operator and the propagator --- which is vital for describing the antiparticles --- it is necessary to use the $\theta$ functions as in \eq{gtheta}. This, in turn, requires the time-ordered correlator, which leads to the right hand side in \eq {qft1134} involving two field operators. 
So the $|t|$, time-ordered correlator and the existence of antiparticles are closely related. 

It is therefore intriguing to see how this $|t|$ arises from the more exact description containing the \zpl. We now have $\sqrt{t^2-L^2}$ (as argued earlier, we will now assume $t^2>L^2$) instead of $|t|$; when we take the limit of $L\to 0$ we get the expression $\sqrt{t^2}$ with two possible signs for the square root. It makes physical sense to define  $\sqrt{t^2}$ as an even function of $t$ by taking:
\begin{equation}
 \sqrt{t^2}=\theta(t)t+\theta(-t)(-t)=|t|
 \label{modt}
\end{equation} 
This will lead to the correct limiting behaviour and standard QFT when $L\to0$, as it should. For $t^2\gg L^2$ we get the expansion:
\begin{equation}
 e^{-i H\sqrt{t^2-L^2}}=e^{-i H|t|}\left[1+\frac{iHL^2}{2|t|} \right]
 =\theta(t)e^{-iHt} \left[1+\frac{iHL^2}{2t} \right]+\theta(-t)e^{iHt}\left[1-\frac{iHL^2}{2t} \right]
\end{equation} 
It is not easy to interpret this cleanly in terms of particle -antiparticle propagation. The result suggests that even the basic notion of particles and antiparticles might require revision close to Planck scales.
This fact is also apparent from the fact the QG corrected propagator \textit{cannot} be expressed as the time-ordered correlator of an underlying quantum field operator. The standard QFT description, when particles emerge as excitations of an underlying operator fails near Planck scales, even though the propagator itself remains well-defined.

\subsection{Speculations about trans-Planckian scales}

The discussion so far is mathematically well-defines and arises as  a direct consequence of our ansatz $\sigma^2\to\sigma^2+L^2$ to capture mesoscopic scale physics. 
Let me now consider the form of the evolution operator for $t^2=(t_2-t_1)^2<L^2$, i.e., at sub-Planckian scales. Conceptually, we \textit{cannot} use our ideas of mesoscopic scales --- and a QG corrected propagator in an effective geometry --- at sub-Planckian scales.
It is however tempting to  speculate as to what the result could mean when $t^2 < L_P^2$. Very often in physics, mathematical structures allow extrapolation of concepts  beyond their originally defined domain of validity thereby leading to fresh insights. With this possibility in mind, I will now speculate as to  what happens to the above results when $t^2 < L_P^2$. 

Let us begin with  Schwinger representation for the QG corrected propagator given in \eq{qgp1} with the $i\epsilon$, $i\delta$ factors explicitly displayed:
\begin{equation}
 G_{QG}(x)= \frac{1}{i} \frac{1}{16\pi^2} \int_0^\infty\frac{ds}{s^2} \, \, \exp -i\left( \frac{x^2-L^2-i\delta}{4 s} +  (m^2-i\epsilon) s\right)
 \label{qggreg}
\end{equation} 
This expression can be integrated exactly as in standard QFT (in the limit of $L=0$) to give the result in \eq{qft176} with $x^2$ replaced $x^2 - L^2$ with $i\epsilon$ prescriptions implicitly understood. This means that the QG corrected propagator, expressed in Schwinger representation in \eq{qggreg}, is well defined for \textit{all} values of $t^2 - |\bm{x}|^2 - L^2$. This is obvious from the fact that the integral in \eq{qggreg} converges for all values of $t^2 - |\bm{x}|^2 - L^2$ because of our $i\epsilon, i\delta $ prescriptions. So, while the expression is \textit{conceptually} meaningful only when $t^2 \gtrsim L^2$ and $|\bm{x}|^2 \gtrsim L^2$, it is \textit{algebraically} meaningful even at sub-Planckian scales; the addition of a zero-point-length merely shifts the location of light cone (where $x^2 =0$) in the spacetime. 

Since the Schwinger representation remains well defined for the QG corrected propagator even at sub-Planckian scales, it is obvious that we should be able to \textit{define} other representations for the propagator as well, for sub-Planckian scales, with suitable choice of square-root conventions etc.  Let us, for example, consider the equivalence between Schwinger representation in \eq{qggreg} and the one in \eq{grb1} which has a square-root, $\sqrt{t^2-L^2}$ in the phase. To check the equivalence of \eq{qggreg} and \eq{grb1} explicitly, we will take the \textit{spatial} Fourier transform of \eq{qggreg}. This requires the computation 
\begin{equation}                                                                                                                                                                                                                                                                                                                                                                                                                                                                                                                                                                                                                               
\int G_{QG}(x_2;x_1) e^{-i\mathbf{p\cdot x}}d^3\mathbf{x} = -\frac{i}{16\pi^2}\int_0^\infty \frac{ds}{s^2} e^{-i m^2s -i[(t^2-L^2)/4s]}
 \int d^3 \mathbf{x}\ e^{i|\mathbf{x}|^2/4s -i\mathbf{p\cdot x}}
\end{equation} 
Evaluating the Gaussian integrals over $\bm{x}$, and writing $s=\rho^2$, we find that:
\begin{equation}
 \int G_{QG}(x_2;x_1) e^{-i\mathbf{p\cdot x}}d^3\mathbf{x} = \left(\frac{i}{\pi}\right)^{1/2} \int_0^\infty d\rho \exp\left( -i\omega^2_\mathbf{p} \rho^2 - \frac{i(t^2-L^2)}{4\rho^2}\right)
\end{equation} 
Recall, \textit{from standard QFT}, that $\omega_{\bf p}^2 $ is actually $\omega_{\bf p}^2 -i\epsilon$ while $t^2 - L^2$ is actually $t^2 - L^2 - i\delta$.  (That is,  we are \textit{not} introducing at this stage any extra prescription and merely using what is required even in the case of standard QFT, corresponding to $L=0$.) To evaluate this integral, we have to use the result
\begin{equation}
 I(a,b) = \int_0^\infty dx\ e^{-i(a-i\epsilon)x^2 -i(b-i\delta)x^{-2}} = \frac{1}{2}\left(\frac{\pi}{ia}\right)^{1/2} \exp\left( - 2i\sqrt{a-i\epsilon}\sqrt{b-i\delta}\right)
\end{equation} 
  This integral is well defined for all real $(a,b)$, positive or negative, because of the $i\epsilon$, $i\delta$ regulators. The result can be easily proved when $a$ and $b$ are positive and the result can be analytically continued for, say, $a>0, b<0$ (which is the case we are interested in) as well. This leads to the result 
  \begin{equation}
 \int G_{QG}(x_2;x_1) e^{-i\mathbf{p\cdot x}}d^3\mathbf{x} = \frac{1}{2\omega_\mathbf{p}}\exp(-i\omega_\mathbf{p} \sqrt{t^2 - L^2-i\delta})
 \label{start11}
  \end{equation}
 (Considering the importance of this result I have provided yet another derivation,  by analytic continuation from the Euclidean sector --- where we do not need the  
  $i\epsilon$, $i\delta$ regulators and integrals are well-defined --- in the Appendix.)
  Inverting the Fourier transform in \eq{start11}, we can write 
  \begin{equation}
  G_{QG}(x_2;x_1) = \int\frac{d^3\mathbf{p}}{(2\pi)^32\omega_\mathbf{p}}  e^{i\mathbf{p\cdot x}} 
e^{-i\omega_\mathbf{p}\sqrt{t^2 - L^2-i\delta}}
\label{start2}
  \end{equation} 
  As we had noted before, the Schwinger representation (and its explicit evaluation in terms of modified Bessel function) tells us  that the left hand side of this equation is well defined. On the right hand side no issues arise when $t^2 > L^2$. When $t^2 < L^2$ the square root has to be defined as $-i|\sqrt{L^2-t^2}|$
  so that the integral is exponentially damped for large values of $|\bm{p}|$.  This is a consistent interpretation of the branch-cut of the square root in complex plane.
  
  The same result can also be obtained (more rigorously) from our result in \eq{defH}. If we replace $|t|$ by $\sqrt{t^2-L^2}$ on both sides, we get the integral representation:
\begin{align}
	U_{QG}(t)=f[H,\sqrt{t^2-L^2}]=e^{-iH\sqrt{t^2-L^2}}
	=\int_{-\infty}^{\infty}\frac{ds}{(i\pi)}\left[\frac{s}{t^2-L^2-s^2-i\epsilon}\right]e^{-iH s}
	\label{defH1}
\end{align}
Since the function $f(H,z)$ is defined everywhere in the  complex plane of $z$, this representation is defined for both signs of $(t^2-L^2)$. Since
$f(H, z=iy)=e^{-H|y|}$ for positive definite $H$ and $y$  real, it follows that $U_{QG}(t)=e^{-H\sqrt{L^2-t^2}}$ for $t^2<L^2$.

Therefore, our result strongly suggests the interpretation of the evolution operator as:
\begin{equation}
G_{\rm QG}(x_2,x_1)  =
\begin{cases}
\bk{\bm{x}_2}{e^{-iH\sqrt{t^2-L^2}}}{\bm{x}_1}; &(\mbox{for}\ t^2 > L^2)\\
        \bk{\bm{x}_2}{e^{-H\sqrt{L^2-t^2}}}{\bm{x}_1}; &(\mbox{for}\ t^2 < L^2)
 \end{cases}
\end{equation} 
Clearly the time evolution operator is not unitary for $|t_2-t_1|<L$, i.e at sub-Planckian scales. 
This is consistent with the idea that the QG-corrected metric could make the spacetime Euclidean at sun-Planckian scales \cite{cheshire, dawood}. Both ideas certainly need to be explored further, checked for inconsistencies etc. I hope to address this question in a future work.

\section*{Acknowledgement}
I thank Sumanta Chakraborty, Dawood Kothawala and Karthik Rajeev for comments on an earlier draft.  My research  is partially supported by the J.C.Bose Fellowship of Department of Science and Technology, Government of India.

\appendix
\section{QFT Propagator without QF}

 The complete dynamics of spinless particle of mass $m$, in a curved spacetime with metric $g_{ab}$ is contained in the  propagator $G_{std}(x_2,x_1)$, or, equivalently, in the rescaled propagator $\mathcal{G} \equiv m G_{\rm std}$. (The latter  will turn out to be simpler to handle algebraically.)\footnote{\textit{Notation:} I will add the subscript `std' for quantities pertaining to a classical gravitational background, not necessarily flat spacetime; the subscript `QG' will give the  corresponding quantities with quantum gravitational correction. For expressions corresponding to a free quantum field in flat spacetime I use the subscript `free'.} I will now introduce three definitions for this  propagator,   which are robust enough to survive (and be useful) at mesoscopic scales. 
 
 All these  three, equivalent, ways of defining this propagator \textit{works without using the notion of a local quantum field operator, canonical quantisation, vacuum state etc.}.\footnote{Doing some reverse-engineering, it is possible to obtain the $G_{QG}$ as a two-point-function of a highly nonlocal field theory; see eq 37 of \cite{mpla20}. But the non-locality of the theory makes it difficult to analyze it along standard lines.} 
 The first definition of the (Euclidean) propagator\footnote{In this Appendix, I will work in a Euclidean space(time) and will assume that the results in spacetime arise through analytic continuation. This is \textit{not} crucial and one could have done everything in the Lorentzian spacetime itself.} is given by: 
 \begin{equation}
\mathcal{G}_{\rm std}(x,y;m)\equiv
mG_{std}(x,y;m^2)=\int_0^\infty m\ ds\ e^{-m^2s}K_{std}(x,y;s)
\label{a14}
 \end{equation} 
 where $K_{\rm std}$ is the zero-mass, Schwinger (heat) kernel given by $K_{\rm std} (x,y;s) \equiv \bk{x}{e^{s\Box_g}}{y}$. Here $\Box_g$ is the Laplacian in the background space(time.
This heat kernel is a purely geometric object,  determined by the background  geometry. It has the form (in $D=4$):
\begin{equation}
 K_{std}(x,y;s)\propto\frac{e^{-\bar\sigma^2(x,y)/4s}}{s^2}\left[1+ \text{curvature corrections}\right]
\end{equation} 
where $\bar\sigma^2(x,y)$ is the geodesic distance. The curvature corrections, encoded in the Schwinger-Dewitt expansion, will involve powers of $(s/L_{curv}^2)$. The exponential $e^{-m^2s}$ in \eq{a14} suppresses the integral for $s\gtrsim \lambda_c^2$  
(where $\lambda_c = \hbar/mc$ is the Compton wavelength of the particle)
and hence, when $\lambda_c\ll L_{curv}$, the curvature corrections will be small.

The second definition of the propagator we can use is based on the path integral sum: 
 \begin{equation}
\mathcal{G}_{\rm std}(x_1,x_2;m) = \sum_{\rm paths\ \sigma} \exp -m\sigma (x_1,x_2)
  \label{three}
 \end{equation}
 where $\sigma(x_1,x_2)$ is the length of the path connecting the two events $x_1, x_2$ and the sum is over all paths connecting these two events.
 This sum can be defined in the lattice and computed --- with suitable measure --- in the limit of zero lattice spacing \cite{pid,tpqft}. The result will, of course, agree with that in \eq{a14}.
 
 The third definition is an interesting variant of this which has not been  explored in the literature. This is  obtained by converting the path integral to an \textit{ordinary} integral. To do this, let us 
 introduce a Dirac delta function into the path integral sum in \eq{three} and use the fact that both  $\ell$ and $\sigma$  are positive definite, to obtain:
 \begin{equation}
\mathcal{G}_{\rm std}(x_1,x_2;m) = \int_{0}^\infty d\ell\ e^{-m\ell} \sum_{\rm paths\ \sigma}\delta_D \left(\ell - \sigma (x_2,x_1)\right)
 \equiv \int_{0}^\infty d\ell\ e^{-m\ell} N_{std}(\ell; x_2,x_1)
  \label{k2}
 \end{equation}
 where I have defined the function $N_{std}(\ell; x_2,x_1)$ to be:
 \begin{equation}
N_{std}(\ell; x_2,x_1) \equiv \sum_{\rm paths\ \sigma}\delta_D \left(\ell - \sigma (x_2,x_1)\right)
  \label{k3}
 \end{equation}
 The last equality in \eq{k2}
 converts the path integral to an \textit{ordinary} integral with a measure $N(\ell)$ which --- according to \eq{k3} --- can be thought of as counting the \textit{effective} number of paths\footnote{The \textit{actual} number of paths, of a specified length, connecting any two points in the Euclidean space, is either zero or infinity. But the \textit{effective} number of paths $N(\ell)$, defined as the inverse Laplace transform of $\mathcal{G}$ (see \eq{k2}), will turn out to be a finite quantity.} of length $\ell$ joining the two events $x_1$ and $x_2$. Usually, I will just write $N(\ell)$ without displaying the dependence on the spacetime coordinates to keep the notation simple. 
 
 Let me illustrate the form of $N(\ell)$ in the case of a free field in flat space. Expressing both $\mathcal{G}_{\rm free}(p,m)=m(p^2+m^2)^{-1}$ and $N_{\rm free}(p,\ell)$ in momentum space, we see that:
 \begin{equation}
  \mathcal{G}_{\rm free}(p^2,m) = m G_{\rm free}(p^2,m^2) = \frac{m}{m^2+p^2} = \int_0^\infty d\ell\ e^{-m \ell} \cos p\ell
   \label{thirteen}
  \end{equation} 
  That is, the $N_{\rm free}(p,\ell)$ in momentum space is given by the simple expression $N_{\rm free}(p, \ell)$  $= \cos(p\ell)$. (The form of $N_{\rm free}(\ell, x_2,x_1)$ in real space can also be computed in closed form by a Fourier transform; see e.g., \cite{ppoe}.)
  
 It is easy to understand how the introduction of \zpl\ into the  geometry modifies the propagator in \eq{k2}. The existence of the \zpl\ suggests that we should change the path length $\ell$ appearing in the amplitude to $(\ell^2 + L^2)^{1/2}$. Therefore the quantum corrected propagator will be given by the last integral in \eq{k2} with this simple replacement. This leads to the expression for the propagator incorporating the \zpl:
 \begin{equation}
  \mathcal{G}_{\rm QG} (x_1,x_2; m) = \int_0^\infty d\ell \ N_{std}(\ell; x_1,x_2) \exp\left( - m \sqrt{\ell^2+L^2}\right)
  \label{b61}
 \end{equation} 
 The  modification $\ell \to (\ell^2 + L^2)^{1/2}$ ensures that all path lengths are bounded from below by the \zpl.\footnote{One can also obtain the same result by modifying $N_{std}$ to  another expression $N_{QG}$ and leaving the amplitudes the same. But the above interpretation is more intuitive.} 
 
 The original path integral in \eq{k2} had an equivalent description in terms of the heat kernel through \eq{a14}. The modification in \eq{b61} translates to a modified relation between the heat kernel and the propagator.  With some elementary algebra, involving Laplace transforms \cite{ppoe}, one can show that \eq{a14} is now replaced by: 
 \begin{equation}
  \mathcal{G}_{QG}(x,y;m)=\int_0^\infty m\ ds\ e^{-m^2s-L^2/4s}K_{std}(s; x,y)
  \label{a151}
 \end{equation} 
 This was the result, \eq{a15}, used in the main text. 
 
 Again, let me illustrate both \eq{b61} and \eq{a151} --- which are actually valid in arbitrary curved spacetime --- in the simple context of a free field in  flat spacetime. In the momentum space we can use the result $N_{free}(p,\ell)=\cos pl$ in \eq{b61}, to get:
 \begin{equation}
  \mathcal{G}_{\rm QG}(p^2) =  \int_0^\infty d\ell \ e^{-m \sqrt{L^2+\ell^2}} \ \cos(p\ell) = \frac{mL}{\sqrt{p^2+ m^2}} K_1[ L\sqrt{p^2+m^2}]
   \label{fifteen1}
  \end{equation} 
 Similarly, using the expression for  zero-mass, flat-space kernel in the momentum space,, $K_{std}(s;p)=\exp(-sp^2)$ in \eq{a151} we find that:
 \begin{equation}
   \mathcal{G}_{\rm QG}(p^2) = \int_0^\infty ds \ m\ \exp\left[-s(p^2+m^2) - \frac{L^2}{4s}\right] = \frac{mL}{\sqrt{p^2+m^2}} \ K_1[L\sqrt{p^2+m^2}]
   \label{start1}
  \end{equation} 
which is identical to \eq{fifteen1}. These expressions describe the QG corrections to the propagator in a freely-falling-frame \cite{ppoe}. 

The propagator with \zpl\ also has an elegant path integral description \cite{pid}. A heuristic way of obtaining this is as follows:  The path integral in  \eq{three} implies that the amplitude  is exponentially suppressed for paths longer than the  Compton wavelength $\lambda_c \equiv \hbar/mc$. This is due to the fact that the action for a relativistic particle of mass $m$ leads to the factor $\exp(-A/\hbar)$ with $A/\hbar = -mc\sigma/\hbar = - \sigma /\lambda_c$ where $\sigma $ is the length of the path and $\lambda_c = \hbar/mc$ is the Compton wavelength of the particle. There is also  another length scale --- viz. the gravitational Schwarzschild radius $\lambda_g \equiv Gm/c^2$ --- which we can associate with a particle of mass $m$. It makes absolutely no sense to sum over paths with $\sigma \lesssim \lambda_g$ in the path integral.  Just as paths with $\sigma \gtrsim \lambda_c$ are suppressed exponentially by the factor $\exp[-(\sigma/\lambda_c)]$, it is necessary to suppress exponentially the paths with $\sigma \lesssim \lambda_g$  by another exponential\footnote{Why this factor should also be exponential, rather than of some other functional form, is a nontrivial question and is closely related to principle of equivalence. It is explained in detail in Ref. \cite{ppoe}.} factor $\exp[-(\lambda_g/\sigma)]$.
So, a natural and minimal modification of the path integral sum in \eq{three},  which incorporates the Schwarzschild radius of a particle of mass $m$, will lead to the path integral sum: 
 \begin{equation}
  \mathcal{G} (x_1,x_2) \equiv \sum_{\rm paths\ \sigma} \exp\left[-\frac{\sigma}{\lambda_c}\right]\ \exp\left[-\frac{\lambda}{\sigma}\right]
  = \sum_{\rm paths\ \sigma} \exp\left[- m \left(\sigma + \frac{L^2}{\sigma}\right)\right]
  \label{pid}
 \end{equation}   
 where $L= \mathcal{O}(1) L_P$.
 This path sum can also be evaluated on a lattice \cite{pid} and leads to the same expression for $G_{QG}$ as the two previous definitions. 
 The path integral, given by \eq{pid} also has a beautiful symmetry: The amplitude is invariant  under the duality transformation $\sigma \to L^2/\sigma$.
 
 \section {Analytic continuation to Lorentzian sector}
 
 In this Appendix, I will briefly outline how the different results in the Lorentzian sector, used in the main text, arises from the analytic continuation from the Euclidean sector. To begin with, let me write write down the Schwinger representation for the propagator in the Euclidean sector by Fourier transforming $G_{QG}(p^2)=\mathcal{G}_{\rm QG}(p^2)/m$ in \eq{start1} with respect to $p$. Evaluating the Gaussian integrals immediately gives:
 \begin{equation}
 G_{\rm QG} (x) = \frac{1}{16\pi^2}  \int_0^\infty \frac{ds}{s^2} \ \exp\left(-sm^2 - \frac{x^2+L^2}{4s}\right)
  \label{X1}
 \end{equation} 
In the Euclidean sector $x^2=t_E^2+|\bm{x}|^2$ which goes over to $-t^2+|\bm{x}|^2=-x^2$ on analytic continuation with our mostly negative signature. Further $s$ is replaced by $is$ (which is easy to see from the fact that $e^{-m^2s}$ should go over to $e^{-im^2s}$). So, analytic continuation of \eq {X1} gives:
\begin{equation}
 G_{\rm QG} (x) = \int_0^\infty\frac{ds}{16\pi^2 i}\  \exp\left(-im^2s - \frac{i}{4s}(x^2-L^2)\right)
\end{equation} 
which, of course, is the same as \eq{qgp1} used in the main text.

Let me now provide another derivation of the result in \eq{start2} by working in the Euclidean sector and analytically continuing the result. Since the spatial coordinates  are unchanged when we go from Euclidean to Lorentzian sector, we can start with the spatial Fourier transform of Euclidean propagator in \eq{X1}. Writing $x^2=t_E^2+|\bm{x}|^2$ in \eq{X1} and evaluating the Gaussian integrals over $\bm{x}$, we get:
\begin{align}
\int G_{\rm QG}(x) \ e^{-i\bm{p\cdot x}} d^3\bm{x} 
&=\frac{1}{\sqrt{4\pi}} \int_0^\infty\frac{ds}{\sqrt{s}}\  \exp\left(-\omega^2_{\bm{p}}s - \frac{t_E^2+L^2}{4s}\right)\nonumber\\
&=\frac{1}{\sqrt{\pi}}\int_0^\infty d\rho\  \exp\left(-\omega^2_{\bm{p}}\,\rho^2 - \frac{t_E^2+L^2}{4\rho^2}\right)
\end{align} 
where $\omega_{\bm{p}}^2=\bm{p}^2+m^2$ and
we have substituted $s=\rho^2$ to get the last expression. This integral, of course is perfectly well defined without requiring any regulators and can be evaluated using the standard result:
\begin{equation}
\int_0^\infty dx\ \exp\left( -A^2 x^2 - \frac{B^2}{x^2}\right) = \frac{1}{2} \frac{\sqrt{\pi}}{|A|} \, \exp\left(-2|A|\, |B|\right)
\end{equation} 
This immediately gives the final answer in the Euclidean sector:
\begin{equation}
 \int G_{\rm QG}(x) \ e^{-i\bm{p\cdot x}} d^3\bm{x} = \frac{1}{2\omega_{\bm{p}}} \ \exp\left( -\omega_{\bm{p}} \sqrt{t_E^2 + L^2}\right) 
\end{equation} 
The analytic continuation to the Lorentzian sector involves the replacement:
\begin{equation}
 \omega_{\bm{p}} \sqrt{t_E^2 + L^2} \to \omega_{\bm{p}} \sqrt{-t^2 + L^2} = i \omega_{\bm{p}} \sqrt{t^2 - L^2}
 \label{X7}
\end{equation} 
which reproduces the result in \eq{start2} without the use of regulators for integrals etc. The sign of the square root in taken to be positive  in \eq{X7} in order to reproduce the standard QFT result when $L=0$, along the lines of \eq{modt}.

Finally, I provide another  integral representation for the time evolution operator $\exp(-iH\sqrt{t^2-L^2})$ from the Fourier space expression for the propagator. To do this we compute the Fourier transform of $\exp(-iH\sqrt{t^2-L^2})$ with respect to $t$ by multiplying both sides of \eq{new117} by $e^{i\nu t}$ and integrating over $t$ along the whole real line.
We find, after some simple algebra that
\begin{align}
\int_{-\infty}^\infty dt\  e^{i\nu t - i H \sqrt{t^2 - L^2}}
 &= 2H \int_0^\infty d \rho\ \exp\left( -i\rho ( H^2 - \nu^2 - i\epsilon) + \frac{i(L^2 - i \delta)}{4\rho}\right)\nonumber\\
 &=\frac{2H}{i} \left[ \frac{L^2}{4(H^2-\nu^2)}\right]^{1/2} \, K_1 \left[ \sqrt{L^2 (H^2 - \nu^2)}\right] \equiv F_\nu(L,H)
\end{align} 
In the first equality we have explicitly displayed the $i\epsilon,i\delta$ factors which ensures convergence. In the final expression it is understood that $H^2=H^2-i\epsilon$ and $L^2=L^2-i\delta$. This allows us to write:
\begin{equation}
 e^{-iH\sqrt{t^2-L^2}}=\int_{-\infty}^{\infty} \frac{d\nu}{2\pi}F_\nu(L,H)e^{-i\nu t}
 \label{k1ft}
\end{equation} 
 It is straightforward to verify that (since $K_1(z)\approx(1/z)$ as $z\to0$), \eq{k1ft} reduces to standard result in the limit of $L\to0$, as it should. This expression, with $H^2$ replaced by $-\omega_{\bm{p}}^2$ also provides an explicit realization of the result mentioned in \eq{ix2}.

\end{document}